# Effect of adaptive cruise control systems on mixed traffic flow near an on-ramp

L. C. Davis*, Physics Dept., University of Michigan, Ann Arbor, MI 48109

**Abstract**

Mixed traffic flow consisting of vehicles equipped with adaptive cruise control (ACC) and manually driven vehicles is analyzed using car-following simulations. Unlike simulations that show suppression of jams due to increased string stability, simulations of merging from an on-ramp onto a freeway have not thus far demonstrated a substantial positive impact of ACC. In this paper cooperative merging is proposed to increase throughput and increase distance traveled in a fixed time (reduce travel times). In such a system an ACC vehicle senses not only the preceding vehicle in the same lane but also the vehicle immediately in front in the opposite lane. Prior to reaching the merge region, the ACC vehicle adjusts its velocity to ensure that a safe gap for merging is obtained. If on-ramp demand is moderate, partial implementation of cooperative merging where only main line ACC vehicles react to an on-ramp vehicle is effective. Significant improvement in throughput (18%) and increases up to 3 km in distance traveled in 500 s are found for 50% ACC mixed flow relative to the flow of all manual vehicles. For large demand, full implementation is required to reduce congestion.

* Email: davislc@umich.edu

# Effect of adaptive cruise control systems on mixed traffic flow near an on-ramp

L. C. Davis, Physics Dept., University of Michigan, Ann Arbor, MI 48109

## 1. Introduction

Adaptive cruise control (ACC) systems are now available on some luxury cars and might be used on a significant fraction of all vehicles in the future. ACC adjusts vehicle speed according to the range and rate of change of range to the preceding vehicle to maintain a safe distance. Throttle control and, in some designs, mild braking can be employed above a cutoff speed to achieve the desired range and velocity. Future generations of ACC could take complete control of the longitudinal motion. For recent discussions of ACC, see Bareket *et al.* (2003) as well as VanderWerf *et al.* (2002).

Several papers (based on simulations) assessing the impact of the increasing proportion of ACC vehicles have appeared. In addition to improving driver comfort and safety, system-wide benefits for freeway traffic have been suggested. Kukuchi *et al.* (2003) as well as Kerner (2003) found that ACC vehicles tend to promote stability of traffic flow. Davis (2004b), like Kerner (2003), showed that ACC can suppress wide moving jams by increasing string stability. Treiber and Helbing (2001) reported that if 20% of vehicles were equipped with ACC, nearly all congestion could be eliminated on a German autobahn. Bose and Ioannou (2003) showed that the flow-density curve for mixed traffic should fall between the curves for manual and ACC vehicles. Ioannou and Stefanovic (2005) also analyzed mixed traffic, considering the effects of unwanted cut-ins (See, for example, Sultan *et al.*, 2002.) due to larger gaps in front of ACC vehicles. They demonstrated that the smoothness of ACC vehicle response attenuates the perturbation due to a cut-in. Overall, Ioannou and Stefanovic (2005) concluded that there were environmental benefits due to reduced exhaust emissions.

Not all the effects of ACC on traffic were found to be favorable. In addition to the cut-in problem, Kerner (2003) found that ACC vehicles might induce congestion at bottlenecks. Only marginal improvements in travel times were found by Bose and Ioannou (2003) and by Davis (2004b) in on-ramp simulations. Since congestion often occurs at bottlenecks caused by merging vehicles, the ability of ACC to alleviate congestion can therefore be questioned. The purpose of the present work is to show how the introduction of simple cooperative merging between ACC vehicles and other vehicles reduces travel delays and increases flow. No attempt is made to describe how the extra capability could be implemented in hardware or software. Only the benefits of such a capability are examined.

This paper is organized as follows. The dynamics of both ACC and manual vehicles is described in Sec. 2. Merging at on-ramps and simulations for mixed traffic without cooperative merging are discussed in Sec. 3. The formalism for cooperative merging is given in Sec. 4. Simulations with cooperative merging are reported in Sec. 5. Conclusions are stated in Sec. 6.

**2. Vehicle dynamics**

*2.1. Adaptive cruise control vehicles*

The dynamics of ACC vehicles for the simulation model is described in this section. For the *n*th vehicle in a lane, the equation of motion is

$$\tau \frac{dv_n(t)}{dt} + v_n(t) = V(\Delta x_n(t), \Delta v_n(t)), \qquad (1)$$

where the distance between the *n*th vehicle and the preceding one (*n-1*) is

$$\Delta x_n(t) = x_{n-1}(t) - x_n(t). \qquad (2)$$

This quantity is the headway (including vehicle length) and its rate of change is the velocity difference

$$\Delta v_n(t) = v_{n-1}(t) - v_n(t). \qquad (3)$$

The mechanical time constant is $\tau$ and the right-hand side of Eq. (1) is

$$V(\Delta x_n(t), \Delta v_n(t)) = \frac{1}{h_d}(\Delta x_n(t) - D + \tau \Delta v_n(t)). \qquad (4)$$

The headway time is $h_d$. Using the work of Liang and Peng (1999, 2000), Davis (2004b) established that this form of control law is string stable.

The constraints imposed by vehicle mechanical limitations on acceleration and deceleration are

$$a_{accel} \geq \frac{dv_n}{dt} \geq -a_{decel}. \tag{5}$$

The maximum acceleration is $a_{accel}$ = 3 m/s² and the maximum deceleration is $a_{decel}$ = 10 m/s². To avoid collisions, the Gipps-like condition (Gipps, 1981)

$$\frac{dv_n}{dt} \leq -a_g \tag{6}$$

must be satisfied when

$$\Delta x_n(t) + \frac{v_{n-1}^2(t) - v_n^2(t)}{2a_g} - t_d v_n(t) < D. \tag{7}$$

Here D = 7 m and $a_g$ is 3 m/s². The condition $dv_n(t)/dt = -a_g$ is imposed when the equation of motion fails to give sufficient deceleration.

*2.2. Manual vehicles*

In addition to the mechanical time constant $\tau$, human drivers do not respond immediately and therefore exhibit a delay time $t_d$ (typically about 0.75 s), which is included in the model. The equation of motion for manual vehicles is given by a generalized Optimal Velocity Model,

$$\tau \frac{dv_n(t)}{dt} + v_n(t) = V_{desired}(t), \tag{8}$$

where

$$V_{desired}(t) = V_{OV}(\Delta_n(t)), \text{ if } V_{OV}(\Delta_n(t)) < v_n(t), \tag{9a}$$

$$\Delta_n(t) = \Delta x_n(t - t_d) + t_d \Delta v_n(t - t_d). \tag{9b}$$

$V_{OV}$ is the optimal velocity function (Bando *et al.*, 1995),

$$V_{OV}(h) = V_0 \{\tanh[C_1(h - h^c)] + C_2\}. \tag{10}$$

To eliminate violent oscillations in velocity and thus stabilize motion it is necessary to replace $V_{OV}$ by the velocity of the preceding vehicle under some conditions (Davis, 2003). For $V_{OV}(\Delta_n(t)) \geq v_n(t)$ the right-hand side of Eq. (9) becomes

$$V_{desired}(t) = \min\{V_{OV}(\Delta_n(t)), \ v_{n-1}(t-t_d)\}, \text{ if } V_{OV}(\Delta_n(t)) \geq v_n(t). \quad (11)$$

Eq. (11) holds if $\Delta_n(t) < 2H_{OV}(v_{n-1}(t-t_d))$, where the inverse function $H_{OV}$ is given by

$$H_{OV}(V_{OV}(h)) = h. \quad (12)$$

In the original Optimal Velocity model of Bando *et al.*, (1995), $H_{OV}$ is the equilibrium headway at a given velocity. For larger headways, the model is further modified to make vehicles catch up. When $\Delta_n(t) \geq 2H_{OV}(v_{n-1}(t-t_d))$ and $V_{OV}(\Delta_n(t)) \geq v_n(t)$, the right-hand side of Eq. (9) becomes

$$V_{desired}(t) = V_{OV}(\Delta_n(t)) + [v_{n-1}(t-t_d) - V_{OV}(\Delta_n(t))]\exp\left(1 - \frac{\Delta_n(t)}{2H_{OV}(v_{n-1}(t-t_d))}\right). \quad (13)$$

The constants in Eq. (10) are given by $C_1 = 0.086$/m, $C_2 = 0.913$, $h^c = 25$ m, and $V_0 = 16.8$ m/s (Sugiyama, 1996).

The generalized Optimal Velocity Model, which is a refinement of the modified Optimal Velocity Model of Davis (2004a), was chosen for these simulations because the dynamical equation of motion, Eq. (8), is similar in form to that for ACC vehicles, Eq. (1). Also, it is one of the few models that is consistent with the requirements of the three-phase model of Kerner (2002). In particular, there exist equilibrium states of motion that occupy a two-dimensional region of the flow-density phase space. The original Optimal Velocity Model of Bando *et al.* (1995) does not have this property because the equilibrium solutions strictly fall on a curve—the fundament diagram. The original model also exhibits unphysical oscillations, especially when a delay time due human reaction time is included, and has a tendency to have collisions. The modifications of Davis (2003) correct these problems.

The mechanical constraints, given by Eqs. (5) and (6), are also imposed as well as the condition Eq. (6) must be satisfied when

$$\Delta x_n(t-t_d) + \frac{v_{n-1}^2(t-t_d) - v_n^2(t-t_d)}{2a_g} - t_d v_n(t-t_d) < D. \tag{14}$$

For either type of vehicle, a speed limit is imposed so that

$$V(\Delta x_n, \Delta v_n) \text{ or } V_{desired}(t) \leq v_{Limit}. \tag{15}$$

If $V(\Delta x_n, \Delta v_n)$ or $V_{desired}$ becomes larger than $v_{Limit}$, it is replaced by the speed limit in the equation of motion. The magnitude of the mechanical time constant $\tau$ is 0.5 to 1.0 s.

## 3. Merging at on-ramps
### 3.1. Rules for merging
In this section, merging of vehicles from an on-ramp (lane 2) into the freeway (lane 1) is described. The same rules for merging are applied to both manual and ACC vehicles. The region for vehicles to merge into lane 1 is of length $d_{merge}$. See Fig. 1. If at time $t$ the vehicle labeled $n$ in lane 2 is selected at random to merge and

$$-d_{merge} < x_n(t-t_d) < 0, \tag{16}$$

$n$ is permitted to change lanes only if the following conditions hold:

$$d_f = x_{nf}(t-t_d) - x_n(t-t_d) > S_f H_{OV}(v_n(t-t_d)) \tag{17}$$

and

$$d_b = x_n(t-t_d) - x_{nb}(t-t_d) > S_f H_{OV}(v_{nb}(t-t_d)) \tag{18}$$

where $nf$ ($nb$) is the vehicle in lane 1 directly in front of (behind) $n$. See Fig. 2. Every 0.05 s on average all vehicles are considered for possible merging. The factor $S_f$ is taken to be 0.7 because this value was found to produce reasonable merging rates that did not interrupt mainline flow substantially.

In simulations the lead vehicle on the on-ramp approaches the downstream end of the merge region as if there is a vehicle at $x = 0$ with $v = v_{Limit}$. Also, if

$$x_n(t - t_d) > -\frac{v_n^2(t - t_d)}{a_g}, \quad (19)$$

the model requires

$$\frac{dv_n(t)}{dt} = -a_g. \quad (20)$$

*3.2. Initial Conditions*

The lead vehicle on lane 1 starts at $x = 0$ at time $t = 0$ and subsequently moves at the speed limit, $x_o(t) = v_{Limit}\, t$. The other vehicles have the initial condition $v_n(0) = V^{initial}$, which is the same for all vehicles and is generally close to the speed limit. The initial positions are given by occupying a fraction $p_1$ of the sites $X(k)$ generated according to

$$X(k) = X(k-1) - h_k \quad (21)$$

where $X(0) = 0$ and $h_k$ is the $k^{th}$ headway selected at random from a distribution $P(h)$ of headways. If all sites are occupied ($p_1 = 1$) then $x_n(0) = X(n)$, $n = 1, 2...$

For $P(h)$ a power-law distribution is a reasonable approximation for the headways observed in freely flowing traffic (Knospe *et al.*, 2002). It can be generated by repeatedly letting

$$\left(\frac{h^0}{h_k}\right)^\mu = r_k, \quad (22)$$

where $r_k$ is a random number $0 \leq r_k \leq 1$ and $h^0$ is the smallest headway. It is determined by the initial velocity $V^{initial}$ according to $h^0 = H_{OV}(V^{initial})$. The average headway is

$$\bar{h} = \frac{\mu}{\mu - 1} h^0. \quad (23)$$

The probability that a headway therefore is in the range $h$ to $h + \delta h$ is $P(h)\,\delta h$ for $h \geq h^0$ where

$$P(h) = \frac{\mu}{h^0}\left(\frac{h^0}{h}\right)^{\mu+1}. \tag{24}$$

A typical value of the power is $\mu = 3$. No headways smaller than $h^0$ are allowed in the initial positions.

The initial positions of vehicles in lane 2 are generated in a similar manner. Sites are occupied with probability $p_2$ and the leading site is offset by $X^{offset}$ to more negative $x$, i.e., $X(0) = -X^{offset}$.

During the initial time interval $0 < t \leq t_d$ all vehicles in each lane advance according to $x_n(t) = x_n(0) + V^{initial}\, t$. From then on, the vehicles move according to the equations of motion Eqs. (1) and (8). Typically simulations involve 600 vehicles and times of 500 s.

*3.3 Simulation Results*

The first simulation is for mixed flow with 50% ACC vehicles randomly dispersed among manually driven vehicles on both lanes. The length of the merge region is $d_{merge} = 300$ m. The mechanical time constant is $\tau = 0.75$ s and the delay time due driver reaction is $t_d = 0.75$ s. The headway time for ACC vehicles is taken to be $h_d = 1.4$ s (VandeerWerf et al., 2002). The initial conditions are determined by letting $h^0 = 50$ m (corresponding to $V^{initial} = 31.6886$ m/s), $n = 3$, $p_1 = 1$, $p_2 = 0.3$, and $X^{offset} = 1000$ m. The speed limit is $v_{Limit} = 32$ m/s.

In Fig. 3, the velocity of vehicles passing $x = -d_{merge}$ is shown as a function time (diamonds). The lower curve pertains to vehicles on lane 1 and the upper curve to on-ramp vehicles. For comparison, the velocities in lane 1 when all vehicles are manual are shown (squares). With 50% ACC vehicles the transition to congested flow occurs later but the velocities are lower.

In Fig. 4, the velocity of vehicles in lane 1 is shown as a function of position at $t = 500$ s for mixed flow (diamonds) and for the flow when all vehicles are manually driven (squares). The spatial extent of the congested region is less in mixed flow than for all manual flow. This somewhat offsets the lower velocities so that the distance traveled in 500 s (Fig. 5) is larger for mixed flow for the vehicles originally on lane 1 (car number 1 – 400) and about the same for those originating on the on-ramp (car number 401 – 600). For mixed (manual) flow, cars 1 to 164 (157) have passed the on-ramp and are freely flowing. The lengthening congested region causes the downward slope of this portion of curves. The difference in distance traveled by all vehicles is shown in Fig. 6. The biggest difference is about 1 km. Distance traveled in a certain time is a more natural output of these simulations than travel time for a certain distance and serves as a surrogate.

**4. Cooperative merging**

Since the mere introduction of ACC vehicles does not appear to reduce congestion significantly near an on-ramp, it is suggested that an addition interaction with an on-ramp vehicle attempting to merge be added. In this section, the way a suitably equipped ACC vehicle can adjust its position to the preceding vehicle in the opposite lane as they approach the merge region is formulated. The objective of cooperative merging is to create a large enough gap so that a merging vehicle can change lanes without slowing down appreciably. Let $z_0$ be a point upstream of the merge region. See Fig. 1. Consider an ACC vehicle in lane 1 with label $n$. Then for $z_0 < x_n(t) < -d_{merge}$ let

$$\alpha = 1 - \frac{x_n(t) + d_{merge}}{z_0 + d_{merge}} \tag{25}$$

and for $-d_{merge} < x_n(t) < 0$

$$\alpha = 1. \tag{26}$$

Let $x^B(t)$ be the position of the nearest preceding vehicle in lane 2. It need not be another ACC vehicle. See Fig. 7. Let

$$V^B(t) = [x^B(t) - x_n(t) - \tau_n(v^B(t) - v_n(t))]/h_{d1}, \qquad (27)$$

where $h_{d1}$ can differ from $h_d$. Only if $V^B(t) < V(\Delta x_n(t), \Delta v_n(t))$, is $V(\Delta x_n(t), \Delta v_n(t))$ replaced by [See Eq. (4).]

$$\widetilde{V}(t) = \alpha V^B(t) + (1-\alpha)V(\Delta x_n(t), \Delta v_n(t)). \qquad (28)$$

This ensures there is a suitable gap on the main line behind the merging vehicle. If $x^B(t) > x_{n-1}(t)$ [$n-1$ is the preceding vehicle in lane 1] then set $\alpha = 0$, so that the main line vehicle does not come too close to the preceding vehicle in the same lane.

For either lane, it is required that

$$\widetilde{V}(t) \leq v_{Limit}. \qquad (29)$$

If $\widetilde{V}(t) > v_{Limit}$ then it is replaced by the speed limit in simulations.

Occasionally simulations revealed a lock-up phenomenon where an on-ramp vehicle becomes stalled at the downstream end of the merge region and a mainline ACC vehicle stops just behind it. In this case, cooperative merging is overridden by setting $\alpha = 0$ in the simulations when the ACC vehicle reaches a small velocity (1-5 m/s) and the lock-up is prevented.

## 5. Simulations with cooperative merging

The effects of cooperative merging are analyzed in this section. Initially, we consider simulations where only ACC vehicles in lane 1 interact with on-ramp vehicles. The parameters for the ACC vehicles are $h_d = 1.4$ s and $h_{d1} = 1.7$ s. The latter is made larger to provide a suitable gap for merging vehicles. The point at which cooperation begins is $z_0 = -1000$ m. The initial conditions and all other parameters are the same as for Figs. 3-6. In Table I, the number of merges in 500 s (M), the number of vehicles to pass $x = 25$ ($Q_3$) [This is just beyond the downstream end of the merge region.], and the difference $Q = Q_3 - M$ are given. If there is no congestion on lane 1, $Q = 198$.

**Table I.**

|            | M  | $Q_3$ | Q   |
|------------|----|-------|-----|
| All ACC    | 64 | 261   | 197 |
| 50% ACC    | 64 | 259   | 195 |
| 30 % ACC   | 64 | 248   | 184 |
| All manual | 64 | 220   | 156 |

Surprisingly the number of merges M remains the same for 0 to 100% ACC vehicles. The increase in the total number of vehicles passing the end of the merge region is 39 (18% increase) for 50% ACC compared to all manual vehicles. Only an insignificant further increase is possible with additional ACC vehicles. The values of $Q$ indicate little or no mainline congestion for 50% ACC. Confirmation of this is shown in Fig. 8. The distance traveled is almost the same for all vehicles for the 50% ACC case. For the last vehicles to emerge from the merge region, the difference between 50% ACC and all manual is about 3 km. Even with just 30% ACC (not shown), there is a substantial difference (as much as 2 km) in the distance traveled relative to all manual vehicles. The cause of the difference is the extensive region of congested flow when there are no ACC vehicles as shown in Fig. 9. Only a small region of decreased speed near the on-ramp occurs with 50% of the vehicles equipped with ACC.

The results shown thus far in this section pertain to ACC vehicles in lane 1 that adjust their headway due interaction with on-ramp vehicles. When on-ramp demand is large it is advantageous to include interactions between ACC vehicles in lane 2 (on-ramp) with vehicles on lane 1. The additional interaction is characterized by

$$V^A(t) = [x^A(t) - x_n(t) - \tau_n(v^A(t) - v_n(t)]/h_{d1}, \qquad (30)$$

where $x^A(t)$ is the position of the nearest preceding vehicle in lane 1. If $V^A(t) < V(\Delta x_n(t), \Delta v_n(t))$, where $V(\Delta x_n(t), \Delta v_n(t))$ is due to the interaction with the preceding vehicle in lane 2, it is replaced by [See Eq. (4).]

$$\tilde{V}(t) = \alpha V^A(t) + (1-\alpha)V(\Delta x_n(t), \Delta v_n(t)). \qquad (31)$$

Setting $p_2 = 0.5$ (compared to 0.3 previously) gives on-ramp demand high enough that 50% ACC vehicles can not suppress the formation of congestion. The velocities of vehicles that pass the entrance to the merge region at $x = -d_{merge}$ are shown in Fig. 10 for all manual (diamonds) and 50% ACC vehicles (squares). Only the ACC vehicles on lane 1 have interaction with the on-ramp vehicles. In each case, congestion sets in at about 100 s and the velocities drop to 5 m/s or less.

If the additional interaction between on-ramp ACC vehicles and vehicles on lane 1 is implemented, the situation improves somewhat as shown in Fig. 11. In this figure the velocity of each vehicle is plotted against its position at $t = 500$ s. The full implementation of cooperative merging (ACC vehicles in both lanes interacting with vehicles in the opposite lane) reduces the region of congestion on lane 1. However, the region of reduced velocity on lane 2 is longer. Yet the throughput is larger for full implementation, 233 compared to 224 vehicles in 500 s. The number of merges is almost the same, 95 and 93 respectively. The net effect, as measured by the distance traveled by all vehicles, is positive as demonstrated in Fig. 12. For full cooperation, the total distance (sum of the individual distances) is $d_{total} = 8.97 \times 10^6$ m; for partial cooperation, $d_{total} =$ 8.76 x $10^6$ m; and for all manual vehicles, $d_{total}$ = 8.56 x $10^6$ m. If all vehicles could move freely at the speed limit, the distance would be $d_{total}$ = 9.60 x $10^6$ m.

**6. Conclusions**

The present work addresses the impact of vehicles with adaptive cruise control on traffic flow with a random mixture of ACC and manually driven vehicles. Previous studies generally showed that stability against the formation of jams could be improved by the addition of ACC vehicles. For example, Davis (2004b) showed that 20% ACC could prevent a transition to the wide moving jam phase. [If one includes limitations on acceleration and deceleration, it takes about 30% ACC to accomplish the same effect.] However, the impact of ACC vehicles on congestion near on-ramps is just as important. In this case, the published literature is not so sanguine.

The simulations (without cooperative merging) reported in present work also showed only modest improvements in throughput with as many as 50% ACC vehicles. The additional distance traveled by vehicles that passed the on-ramp in the first 500 s was less than one km. For vehicles merging from the on-ramp, the distance traveled was a few hundred meters less for some vehicles compared to an all-manual scenario, although the total distance traveled by all vehicles was greater.

To make the introduction of ACC vehicles more effective, the implementation of cooperative merging was suggested. No attempt was made to discuss how such an improved functionality could actually be implemented; only the potential improvement in traffic flow was demonstrated if such a system were feasible. The first simulations were done for partial cooperative merging where only ACC vehicles on the main line interact with (sense) on-ramp vehicles (of both types) and adjust their speed. The flow incoming on the main line was 198 in 500 s (equivalent to 1426 vehicles/h) on a single-lane highway and the on-ramp demand was 64 in 500 s (equivalent to 461 vehicles/h). With 50% ACC, congestion was essentially eliminated and travel distances were almost at free-flow levels. The throughput improved by 18% compared to simulations with all manually driven vehicles. With just 30% ACC, the throughput improved by 13%.

If the on-ramp demand were significantly larger, partial cooperative merging was found to be ineffective and full implementation was required to reduce congestion. For on-ramp demand of 106 vehicles in 500 s (equivalent to 763 vehicles/h), the throughput with 50% ACC was improved approximately 4% when ACC vehicles on both lanes (main line and on-ramp) interacted with vehicles in the opposite lane. The total distance traveled during the first 500 s by the 600 vehicles (of which 233 passed the on-ramp) in the simulation improved by nearly 5%. Thus ACC vehicles were able to reduce the amount of congestion, but even with full cooperative merging did not eliminate it. However, the total demand of (main line and on-ramp) is nearly equal to the 2200 vehicles/h capacity of an all-ACC system with $h_d$ = 1.4 s and a speed limit of 32 m/s. So, it is not surprising that only modest improvements were possible.

The main conclusion to be drawn from this work is that the introduction of cooperative merging, if it could be made feasible, would enable ACC vehicles in mixed traffic to significantly impact flow. Although the improvements might not be as striking as the effects of increased string stability on jamming, they are nonetheless significant and should be considered further.

**Figure Captions**

Fig. 1. Schematic diagram of on-ramp (lane 2) and main line (lane 1). The length of the merge region is $d_{merge}$. The point $z_0$ is discussed in Fig. 7.

Fig. 2. Diagram for illustrating rules for merging. Velocity-dependent safe distances $d_f$ and $d_b$ in front of and behind vehicle *n* are required for it to merge.

Fig. 3. Velocity *vs.* time for vehicles passing entrance to merge region at $x = -d_{merge}$. The diamonds are for mixed flow with 50% ACC vehicles. The upper data are from on-ramp (lane 2) vehicles and the lower data are from the main line (lane 1). The squares are for main line vehicles when all vehicles are manually driven. Here a twenty-car average of the latter is depicted because of considerable scatter in the data. The incoming flow on lane 1 is 198 vehicles in 500 s (equivalent to 1426 vehicles/h) and the on-ramp demand is 64 vehicles in 500 s (equivalent to 461 vehicles/h). The merge region is of length $d_{merge} = 300$ m.

Fig. 4. Velocity *vs.* position at 500 s for vehicles on lane 1. The diamonds are for mixed flow with 50% ACC vehicles and the squares are for flow with all manual vehicles.

Fig. 5. The distance traveled in 500 s as a function of car number. Cars 1-400 originate on lane 1 and 401-600 on lane 2. Flow with 50% ACC vehicles is denoted by diamonds and all-manual flow by squares. Distances are less for those vehicles that traveled through the congested region near the on-ramp.

Fig. 6. The difference (between 50% ACC and all manual flow) in distance traveled in 500 s as a function of car number.

Fig. 7. Diagram for cooperative merging by an ACC vehicle (*n*) on the main line interacting with the preceding vehicle in the same lane (*n-1*) and a vehicle (either ACC or manual) on the on-ramp at $x^B$. ACC vehicles begin to adjust their velocity to open a suitable gap for merging at $z_0$, which is upstream of $x = -d_{merge}$. See Fig. 1.

Fig. 8. Distance traveled in 500 s as a function of car number for ACC vehicles in lane 1 equipped with cooperative merging capability. The parameters are the same as Figs. 3-6. The distances for 50% ACC flow (squares) are nearly the same as for free flow. Velocity adjustment for cooperative merging begins at $z_0 = -1000$ m. Lane 1 demand is 198 vehicles in 500s (equivalent to 1426 vehicles/h) and lane 2 demand is 64 in 500s (equivalent to 461 vehicles/h).

Fig. 9. Velocity *vs.* position at 500 s for vehicles in lane 1. Diamonds denote all manual vehicle flow and squares denote 50% ACC mixed flow. ACC vehicles in lane 1 have cooperative merging capability. Same initial conditions as Fig. 8.

Fig. 10. Velocity as a function of time at $x = -d_{merge}$ for on-ramp demand of 106 vehicles in 500 s (equivalent to 763 vehicles/h). Flow with all manual vehicles is denoted by diamonds and flow with 50% ACC vehicles by squares. ACC vehicles in lane 1 have cooperative merging capability. The lower data are from vehicles on lane 1.

Fig. 11. Velocity *vs.* position at $t = 500$ s for lane 1 and lane 2 vehicles and 50% ACC mixed flow. The diamonds denote full cooperative merging (all ACC vehicles have capability) and the squares denote partial cooperative merging (only ACC vehicles in lane 1 have capability).

Fig. 12. Distance traveled in 500 s *vs.* car number for 50% ACC mixed flow with full cooperative merging (diamonds) and partial cooperative merging (squares) as well as all-manual flow (triangles).

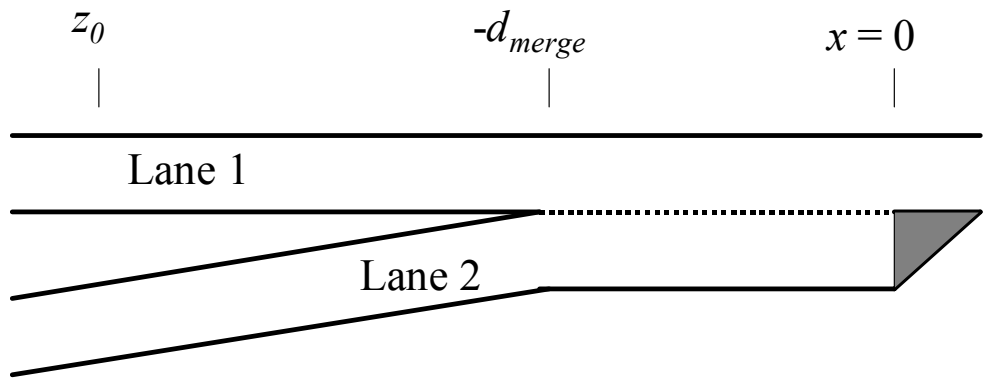

Fig. 1

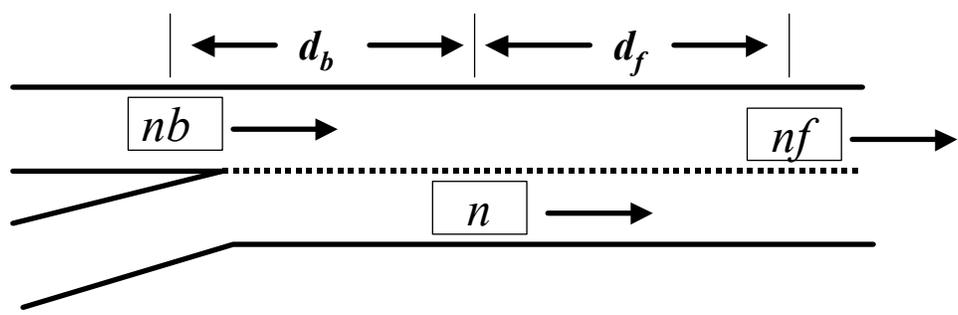

Fig. 2

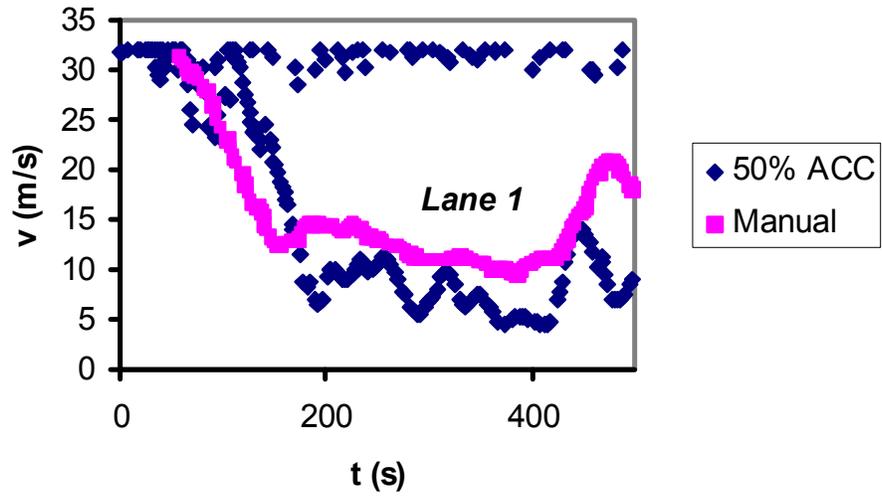

Fig. 3

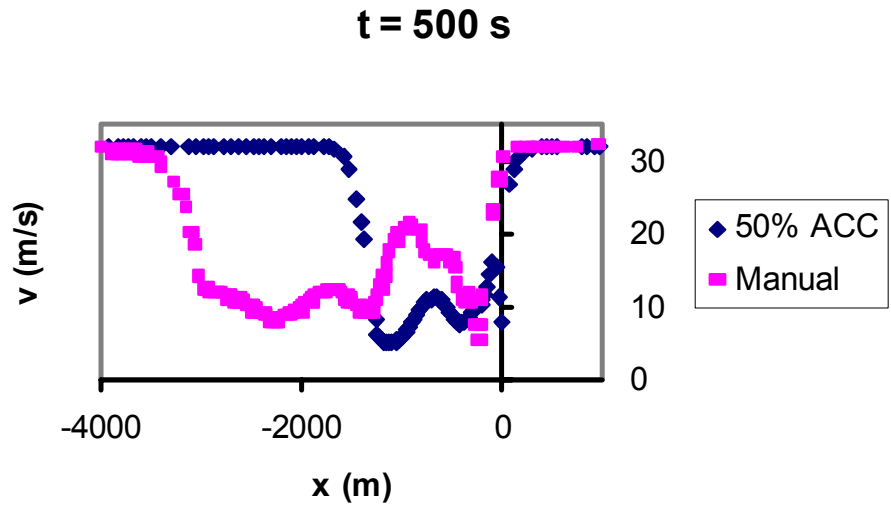

Fig. 4

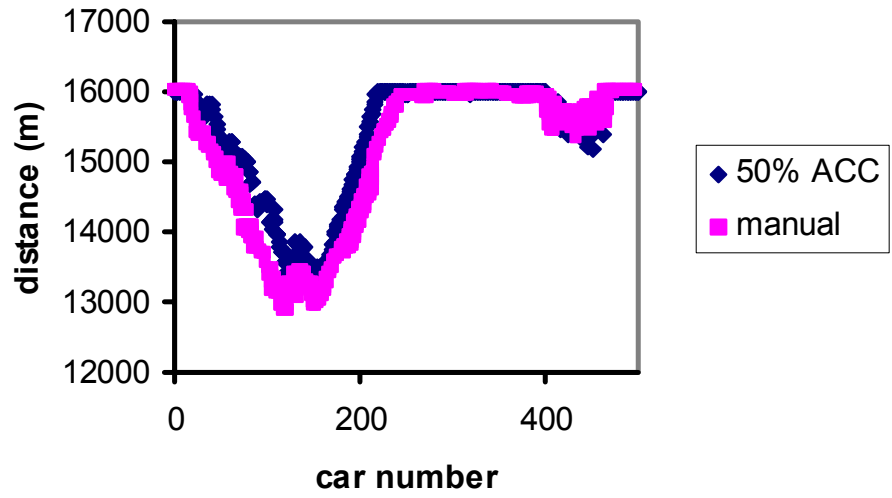

Fig. 5

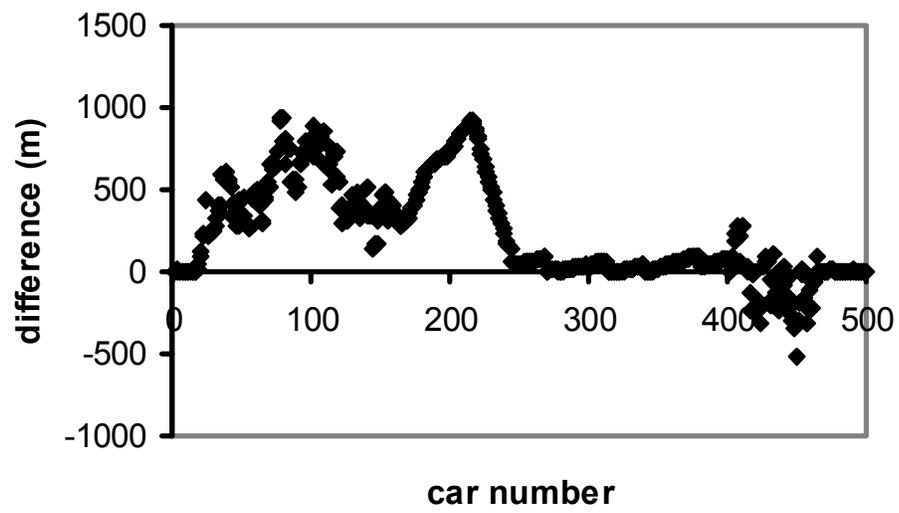

Fig. 6

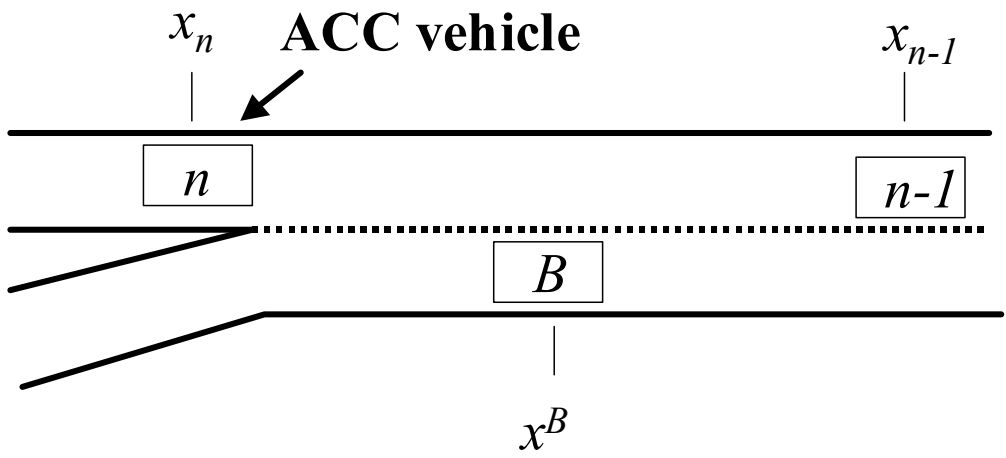

Fig. 7

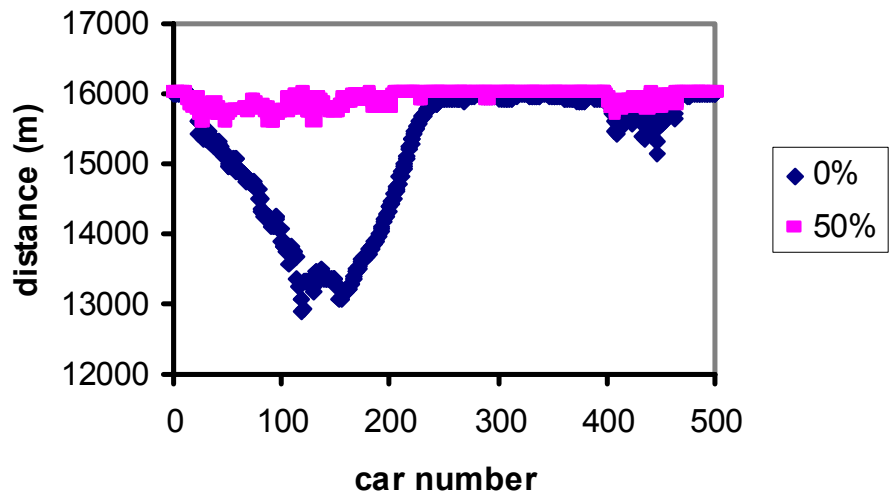

Fig. 8

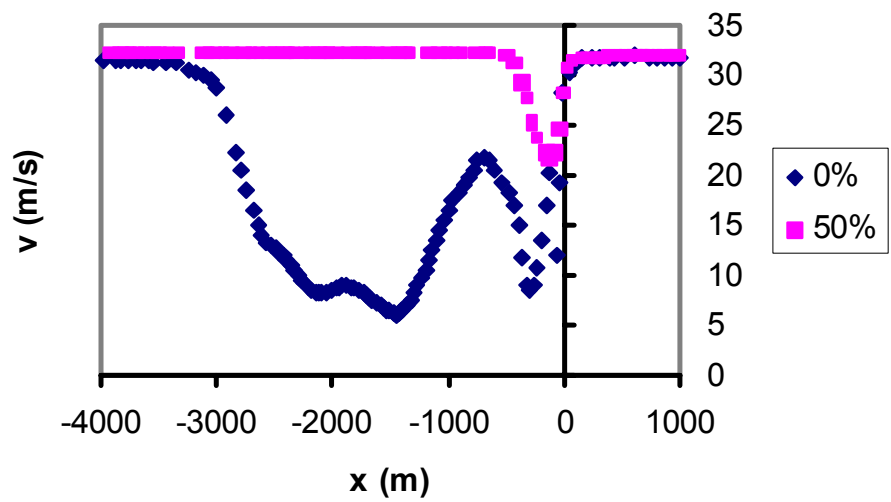

Fig. 9

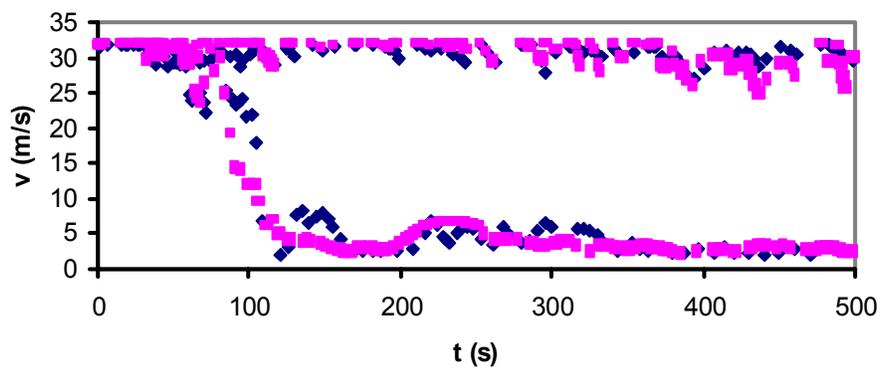

Fig. 10

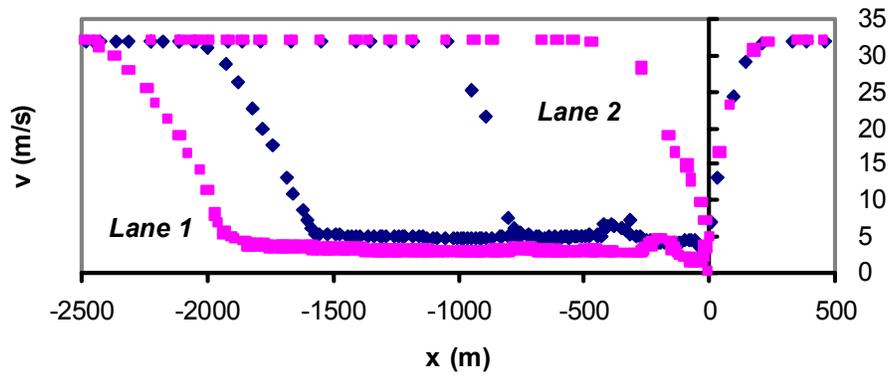

Fig. 11

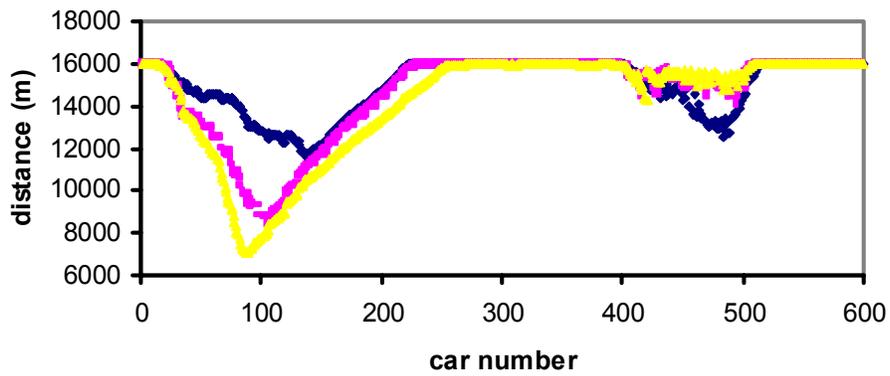

Fig. 12